\documentclass{article}
\usepackage{amsmath, amssymb}
\usepackage[small]{caption}
\usepackage{graphicx,dblfloatfix}
\usepackage{geometry}
\usepackage{hyperref}
\usepackage{mathtools}
\usepackage{physics}
\usepackage{relsize}
\usepackage{subcaption}
\usepackage[inline]{enumitem}
\usepackage{authblk}
\usepackage{cite}

\graphicspath{{./figs/}{.}}

\newcommand*\eq[1]{{#1}^{\left(\mathrm{eq}\right)}}

\begin{document}
	\title{A Partial Entropic Lattice Boltzmann MHD Simulation of the Orszag-Tang Vortex}
	\author{Christopher Flint}
	\author{George Vahala}
	\affil{Department of Physics, College of William \& Mary, Williamsburg, Virginia 23185}
	\date{\today}
	
	\maketitle
	\begin{abstract}
Karlin has introduced an analytically-determined entropic lattice Boltzmann algorithm for Navier-Stokes turbulence.  Here this is partially extended to a lattice Boltzmann model of magnetohydrodynamics, on using the vector distribution function approach of Dellar for the magnetic field. (which is permitted to have field reversal).  The partial entropic algorithm is benchmarked successfully against standard simulations of the Orszag-Tang vortex \cite{Orszag}.  		
	\end{abstract}
	\section{Introduction}\label{Sec:Intro}

Because of its inherent simplicity and ease of being extremely parallizeable, the lattice Boltzmann (LB) algorithm has had considerable impact as a computational tool in the solution of Navier-Stokes flows  \cite{succiBook}.   Interestingly, even though it is strictly a second-order accurate algorithm, in many applications its accuracy is usually comparable to the pseudo-spectral method.  The Achilles' heel of simple LB is that it is prone to numerical instabilities for high Reynolds turbulent flows.  This is attributable to the lack of an inherent mechanism to require the time evolution of the LB distribution function to remain non-negative.

    By generalizing the simple single-relaxation-time (SRT) LB collision operator to incorporate multi-relaxation rates (MRT)  \cite{MRT11,MRT12,MRT13}, one gained some new degrees of freedom that could be exploited to attain greater numerical instability.  However there was no systematic way to constrain these extra relaxation rates, but their tuning was not only problem dependent but could also influence the viscsoity of the flow and hence the actual Reynolds number.  An alternate and systematic approach to stabilizing LB is through an entropic principle and  its assocaited H-theorem.  \cite{entropic1,entropic2,karlinIntro2,boghosianIntro1,boghosianIntro2,boghosianIntro3,keatingIntro1,keatingIntro2,flint9bit,flintMRT,DellarBulkViscosity,DellarMRT,DellarMoments,premnath2009,Bosch,pattison,riley}.  In its current state, the entropic approach can be viewed as an optimized MRT algorithm in which emphasis is placed on an algebraically determined entropy stabilizing parameter that is not directly dependent on the MRT collisional rates and which does not affect the fluid viscosity.  Hence these entropic simulations will be run at the same Reynolds number as the normal CFD computation, rather than at an augmented transport coefficient.
    
    In generalizing the Karlin entropic algorithm to LB-MHD one must decide on the importance of avoiding divergence cleaning (by using a vector magnetic distribution \cite{DellarMHDLB}) over a scalar magnetic distribution that would permit magnetic field reversal.  Here we choose to continue with the vector magnetic distribution representation in which the zeroth moment will yield the magnetic field $\vec B$.  We are thus excluded from applying an extend  entropic principle to the evolution of the vector distribution function.  However there is some residual stabilization effects on the LB-MHD algorithm since the magnetic field appears in the fluid momentum equation to which there is applied an entropic ansatz.  It should also be noted that this entropic stabilization parameter will be dynamically determined analytically throughout the simulation whereas the MRT relaxation rates are static and are not changed throughout the simulation.  Moreover the entropic algorithm will permit simulations at arbitrary small viscosities - a regime where the static MRT algorithms cannot attain.
    
A moment-based representation for LB-MHD is given in Section \ref{Sec:Moments}, while the partial entropic algorithm is presented in Sec. \ref{Sec:Entropic}.  Since 2D and 3D MHD exhibit the same cascading spectra (e.g., energy cascades to high wavenumbers), it is convenient to test our partial entropic LB-MHD algorithm for the 2D Orszag-Teng vortex. Our simulations are compared to those of Orszag-Tang in Sec. \ref{Sec:Simulation}.
	
	\section{Moment Basis Representation for Multiple Relaxation Model of LB-MHD}\label{Sec:Moments}
   There are several MRT extensions \cite{flint9bit,DellarMoments,pattison,Dellar2011} of the original SRT LB-MHD model of Dellar \cite{DellarMHDLB}.  However, for simplicity, we shall restrict ourselves to an SRT model for the vector magnetic field distribution $\vec g_i$, and an MRT model for the scalar distribution function $f_i$
	\begin{eqnarray}
	& \label{LBKinEqn}\left( \partial_t + \partial_\gamma c_{\gamma i} \right) f_i = \sum_j X^{'}_{ij} \left( \eq f_j - f_j \right)  \\
	& \label{LBMagEqn}\left( \partial_t + \partial_\gamma c_{\gamma i} \right) \vec g_i = Y^{'} \left( \vec g_i^{\,(\mathrm{eq})} - \vec g_i \right)
	\end{eqnarray}
We employ the summation convention over the Greek indices which represent the vector nature of the fields ($\gamma=1,2$ for 2D), while the summation over the Roman indices $i, (i = 0 ... 8)$ for a 2D lattice, will always be made explicit.   The relevant moments are 
	\begin{gather}
	\begin{array}{ccccc}
	\sum_i f_i = \rho &,&  \sum_i f_i  \vec c_i = \rho \vec u \quad &\mathrm{and} & \sum_i \vec g_i = \vec B
	\end{array}
	\end{gather}
and the positive definiteness of the $\vec g_i$ is lost for magnetic field reversal problems and thus precludes an explicit attempt at an entropy principle for $\vec g_i$.
 $X^{'}_{ij}$ is the MRT collision operator for the evolution of $f_i$ while $Y^{'}$ is the SRT for the evolution of $\vec g_i$.  The MHD viscosity and resistivity transport coefficients are determined from some of these kinetic relaxation rates.
	
	Since the mean velocity $\vec u$ is defined by the 1st moment of the $f_i$, while the magnetic field $\vec B$ is determined from the 0th moment of the $\vec g_i$, a minimal LB representation of MHD on a square lattice requires a 9-bit model for the $f_i$ but just a 5-bit model for the $\vec g_i$.  However we find it more convenient to use a 9-bit lattice for both distributions, Fig.1.

To recover the MHD equations in the Chapman-Enskog limit, an appropriate choice of relaxation distribution functions  $f{_i}^{(eq)}$ and $\vec g_i^{\,(eq)}$ is (where $w_i$ are appropriate weights, given in Fig. 1)
	\begin{eqnarray}
	&\label{feq} \eq f_i = w_i \rho \left[ 1 + 3\left( \vec c_i \cdot \vec u \right) + \frac{9}{2}\left( \vec c_i \cdot \vec u \right)^2 - \frac{3}{2} \vec u^{\,2} \right] + \frac{9}{2} w_i \left[ \frac{1}{2} \vec B^2 \vec c_i^{\,2} - \left( \vec B \cdot \vec c_i \right)^2 \right]  , i = 0, .. ,8 \\
	& \vec g_i^{\,(\mathrm{eq})} = w_i \!\left[ \vec B + 3 \left\lbrace \left( \vec c_i \cdot \vec u \right) \vec B - \left( \vec c_i \cdot \vec B \right) \vec u \right\rbrace \right] , i = 0, .., 8
	\end{eqnarray}

	\begin{figure}[h]
		\centering
		\includegraphics[width=2.25in, keepaspectratio=true]{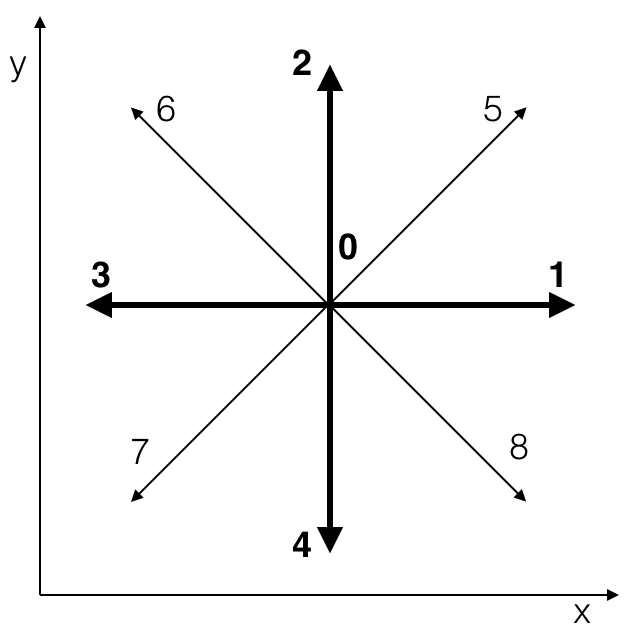}
		\caption[.]{\label{LatticeRep} The kinetic lattice vectors for 2D LB-MHD in $D2Q9$ :
			  $\vec c_i = \left(0,0\right),\left(0,\pm 1\right),\left(\pm 1,0 \right) \ ,\left(\pm 1,\pm 1\right) \ $.  $w_i$ are appropriate weight factors dependent on the choice of lattice: for $D2Q9$ ,$w_0 = \frac{4}{9}$; for speed 1, $w_i = \frac{1}{9}$; and for speed  $\sqrt 2$ , $w_i = \frac{1}{36}$
			.}
		
	\end{figure}
	The LB-MHD equations are typically solved by an operator-splitting method: streaming and local collisional relaxation.  
	The excellent parallelization follows from this: the streaming operation is a simple shift of the data from one lattice point to another, while the collision step is a purely local operation with its evaluation requires only data from only that grid point.  MPI is only required when the data is streamed from one processor domain to another -- and this is well parallelized since it is also synchronized throughout the lattice boundaries.
	The computationally difficult nonlinear $convective$ derivatives $\vec u \cdot \nabla \vec u$, $\vec u \cdot \nabla \vec B$, $\vec B \cdot \nabla \vec u$ and $\vec B\cdot \nabla \vec B$ are replaced in LB by simple linear $advection$ and  local polynomial nonlinearities in $f{_i}^{(eq)}(\vec u, \vec B)$ and $\vec g_k^{\,(eq)}(\vec u, \vec B)$.

Since the collisional relaxation involves the local conservation of mass and momentum at each lattice site, it is convenient in MRT-LB to perform the collision step in moment space.  The streaming is easiest to perform in distribution space 	$f_i, \vec g_i$.  Thus for the moment basis one obviously chooses the conservation moments (the 0th and 1st moments of the $f_i$ and the 0th moment of $\vec g_i$).  The choice of the remaining higher moments is somewhat arbitrary \cite{lallemand1,lallemand2}.  There is a 1-1 (constant) transformation $\mathrm T$ between these spaces.
We choose the same constant $9 \cross 9$  $\mathrm T$- matrix to connect the scalar distributions $(f_i, i=0..8)$ to their moments $(M_i, i=0 .. 8)$ as for the vector magnetic distributions $(\vec g_i, i = 0 .. 8)$ with their moments $(\vec N_i, i = 0 .. 8)$
	\begin{equation}\label{Transformation}
	\begin{array}{ccc}
	M_i = \sum_{j=0}^8 \mathrm T_{ij}f_j &, & \vec N_{i} = \sum_{q=0}^8  \mathrm T_{iq} \vec g_{q}
	\end{array}
	\end{equation}
	where 
	\begin{equation}\label{FluidTMatrix}
	\mathrm {T}=\left( \def\arraystretch{1.2} \begin{array}{c}
	\boldsymbol{1} \\ 
	c_x \\ 
	c_y \\ 
	c_xc_y \\ 
	c^2_x \\ 
	c^2_y \\ 
	c_x^2 c_y \\ 
	c_x c_y^2\\ 
	c_x^2 c_y^2 \end{array}
	\right)=\left( \def\arraystretch{1.2} \begin{array}{rrrrrrrrr}
	1 & \phantom{-}1 & \phantom{-}1 & 1 & 1 & \phantom{-}1 & 1 & 1 & 1 \\ 
	0 & 1 & 0 & -1 & 0 & 1 & -1 & -1 & 1 \\ 
	0 & 0 & 1 & 0 & -1 & 1 & 1 & -1 & -1 \\ 
	0 & 0 & 0 & 0 & 0 & 1 & -1 & 1 & -1 \\ 
	0 & 1 & 0 & 1 & 0 & 1 & 1 & 1 & 1 \\ 
	0 & 0 & 1 & 0 & 1 & 1 & 1 & 1 & 1 \\ 
	0 & 0 & 0 & 0 & 0 & 1 & 1 & -1 & -1 \\ 
	0 & 0 & 0 & 0 & 0 & 1 & -1 & -1 & 1 \\ 
	0 & 0 & 0 & 0 & 0 & 1 & 1 & 1 & 1 \end{array}
	\right)
	\end{equation}             
	The Cartesian components of the corresponding $9$-dimensional lattice vectors are just
	\begin{equation}
	\begin{array}{cccc}
	c_x=\left\{0,1,0,-1,0,1,-1,-1,1\right\} &, &c_y=\{0,0,1,0,-1,1,1,-1,-1\} &.
	\end{array}
	\end{equation}                              

	For the scalar distributions, the 1\textsuperscript{st} row of the $\mathrm T$-matrix is just the conservation of density while the 2\textsuperscript{nd} and 3\textsuperscript{rd} rows are just the conservation of momentum (2D).  For the vector magnetic distributions the 1\textsuperscript{st} row of the $\mathrm T$-matrix is the only collisional invariant.
	
	With this moment basis, the MRT collisional relaxation rate tensor $X^{'}_{ij}$ is diagonalized with the $T-$ matrix as a similarity transformation.  It is convenient to denote this diagonal matrix with elements $X_{i} \delta_{ij}$.

In the $D2Q9$ phase space, the relaxation rate $X_j$ is associated with the corresponding moment $M_j$, $j=0..8$.  Similarly for the magnetic distributions in SRT, there is just a single collisional relaxation rate for each magnetic moment $\vec N_k$, and this will be denoted by  $Y$. 
	In particular, the equilibrium moments can be written in terms of the conserved moments:
	\begin{align}\label{KinMomentEq}
	\begin{split}
	\arraycolsep=15pt
	\begin{array}{lll}
	\eq M_0 = M_0 = \rho & \eq M_1 = M_1 = \rho u_x & \eq M_2 = M_2 = \rho u_y 
	\end{array}
	\\
	\arraycolsep=10pt
	\begin{array}{ll}
	\eq M_3 = \rho u_x u_y - B_x B_y &
	\eq M_4 = \frac{1}{6} \left( 6 \rho u_x^2 + 2 \rho - 3 \left( B^2_x - B^2_y \right) \right) \\
	\eq M_5 = \frac{1}{6} \left( 6 \rho u_y^2 + 2 \rho + 3 \left( B^2_x - B^2_y \right) \right) &
	\eq M_6 = \frac{1}{3} \rho u_y \\
	\eq M_7 = \frac{1}{3} \rho u_x &
	\eq M_8 = \frac{1}{9} \rho \left( 1 + 3 u_x^2 + 3 u_y^2 \right) \\
	\end{array}
	\end{split}
	\end{align}
	\begin{equation}\label{MagMomentEq}
	\arraycolsep=5pt
	\begin{array}{lll}
	\eq N_{\alpha 0} = N_{\alpha 0} = B_\alpha & 
	\eq N_{\alpha 1} = u_x B_\alpha - u_\alpha B_x &
	\eq N_{\alpha 2} = u_y B_\alpha - u_\alpha B_y \\
	\eq N_{\alpha 3} = 0 &
	\eq N_{\alpha 4} = \frac{B_\alpha}{3} &
	\eq N_{\alpha 5} = \frac{B_\alpha}{3} \\
	\eq N_{\alpha 6} = \frac{1}{3} \left( u_y B_\alpha - u_\alpha B_y \right) &
	\eq N_{\alpha 7} = \frac{1}{3} \left( u_x B_\alpha - u_\alpha B_x \right) &
	\eq N_{\alpha 8} = \frac{B_\alpha}{9}, \qquad \alpha = x, y
	\end{array}
	\end{equation}
	 
	\section{Entropic method [6-15] and its partial extension to MHD }\label{Sec:Entropic}
	Initially the entropic LB scheme for Navier-Stokes flows worked with the full distribution function $f_i$ [9-13].  This led to a discrete H-theorem and the need to determine at each lattice site and for every time step the solution of a Newton-Raphson iterative procedure so that one remains on a constant entropy surface.  This is computationally expensive [12, 13] for 3D turbulent flows.  Recently the Karlin group  \cite{entropic1,entropic2} separated the scalar lattice Boltzmann distribution into various moment-related groups and sort an algebraically based entropic parameter for the moment groups above the conserved and shear/stress moments. 	In particular, they formed 3 subgroups for the full $f_i$
	\begin{equation}\label{fExpansion}
	f_i = k_i + s_i + h_i   \quad , \quad i=0 .. 8   .
	\end{equation}
	Here the $k_i$ distributions correspond to the conserved moments, the $s_i$ distributions correspond to the stress/shear moments, and finally the $h_i$ distributions correspond to the remaining higher order moments. 
	
	 For the $k_i$ distributions       
	\begin{equation}\label{contributionExample}
	k_i = \sum_{j=0}^8 \sum_{m=0}^2 \mathrm{T}^{-1}_{im} \mathrm{T}_{mj} f_j   \quad , \quad i=0 .. 8
	\end{equation}
	since there are 3 conserved moments, so the $m-$summation runs from $m=0,1,2$.
	Now the $s_i$ distributions corresponding to the stress/shear moments will come from the deviatoric stress, the trace of the stress tensor and 3rd order moments.  We are at liberty to choose the building blocks of the $s_i$.  Here we let the building blocks be the deviatoric stress and the trace of the stress tensor
		\begin{equation}\label{contributionExample1}
	s_i = \sum_{j=0}^8 \sum_{m=3}^5 \mathrm{T}^{-1}_{im} \mathrm{T}_{mj} f_j   \quad , \quad i=0 .. 8  .
	\end{equation}
where the moments m = 3, 4, 5 are each 2nd order moments in the D2Q9 model.  Finally the remaining moments m = 6, 7, 8 are covered by the $h_i$
  \begin{equation}\label{contributionExample2}
	h_i = \sum_{j=0}^8 \sum_{m=6}^8 \mathrm{T}^{-1}_{im} \mathrm{T}_{mj} f_j   \quad , \quad i=0 .. 8  .
	\end{equation}
	
  A tunable parameter $\gamma$ is introduced to replace the relaxation rates for the higher order subgroup $h_i$, relaxation rates that do not affect the transport coefficients under Chapman-Enskog expansions  \cite{Pope}.  In particular, instead of the standard LB post-collision distributions
	\begin{equation}
	f_i^{'} \equiv f_i\left(t+1\right) = f_i + 2\beta \left(\eq f_i - f_i \right)
	\end{equation}
	\begin{equation}\label{postcollisiongamma}
	\text{we consider} \qquad f_i^{'} = f_i - 2\beta \Delta s_i - \beta \gamma \Delta h_i  .
	\end{equation}
	 $\beta$ is related to the kinematic viscosity : $\nu = \frac{1}{6} \left( \frac{1}{\beta} - 1 \right)$ and we let $\Delta s_i = s_i - \eq s_i$ and $\Delta h_i = h_i - \eq h_i$.  For the conserved moments, of course,  $\Delta k_i = k_i - \eq k_i = 0$.
	
	We wish to maximize the entropy $S \left[ f \right] $ 
	\begin{equation}
	S \left[ f \right] = - \sum_i f_i \ln(\frac{f_i}{w_i}).
	\end{equation}  
	To do this, the entropy is written in terms of the post-collisional state and the $\gamma$ parameter.  The critical point of the entropy \cite{entropic1,entropic2} determines the tunable parameter $\gamma$ from 
	\begin{equation}\label{CritPoint}
	\sum_i \Delta h_i \log(1 + \frac{\left( 1 - \beta \gamma \right) \Delta h_i - \left( 2\beta - 1 \right) \Delta s_i}{\eq f_i}) = 0
	\end{equation}

	Unfortunately, it is a rather computationally expensive root-finding procedure to determine $\gamma(\vec x, t)$ at every lattice grid point at every LB time step.  However, Karlin \textit{et. al }\cite{entropic1,entropic2} noted
	that if one invokes the simple small argument expansion  $\log(1+x)=x+...$  then the entropic factor can be determined algebraically.  Since this is an approximation, we shall denote this entropic factor by $\gamma^{*}$, with
	\begin{equation}\label{gammaSolved}
	\gamma^{*} = \frac{1}{\beta} - \left( 2 - \frac{1}{\beta} \right) \frac{\left\langle \Delta s | \Delta h \right\rangle }{\left\langle \Delta h | \Delta h \right\rangle }
	\end{equation}
	\begin{equation}
	\text{where the inner product}\quad \left\langle A | B \right\rangle = \sum_i \frac{A_i B_i}{\eq f_i}.
	\end{equation}
	Using this $\gamma^{*}$ in the new post-collisional state (Eq. \ref{postcollisiongamma}) one has achieved a maximal entropy state.  The Karlin group have successfully benchmarked this approximation for the tunable parameter $\gamma^{*}(\vec x,t)$ in various 2D and 3D Navier-Stokes simulations \cite{entropic1,entropic2}.  One way to view this entropic algorithm is to consider it a dynamical subset of MRT -  dynamical since the entropic parameter is tuned at every lattice point and every time step algebraically  as opposed to the static relaxation times of a typical MRT simulation.
	
	Clearly, this analysis can not simply carry over to LB-MHD with possible non-positive vector magnetic distributions for magnetic field reversals.  Hence we make the ansatz for our partial entropic algorithm that the entropic parameter in LB-MHD will still be determined by Eq. (\ref{gammaSolved}) for the corresponding LB-MHD $\Delta h$ and $\Delta s$.  The validity of our ansatz will now be tested against the Orszag-Tang vortex simulations \cite{Orszag}.
	
	Thus our partial entropic LB-MHD algorithm consists of the following steps (c.f.,  Karlin \textit{et. al.} \cite{entropic1}:
	\begin{enumerate}
		\itemsep 0em
		\item Compute the conserved moments ($\rho$,$\mathbf u$,$\mathbf B$) (Eq. \ref{Transformation}, \ref{KinMomentEq}, \ref{MagMomentEq})
		\item Evaluate the equilibria $\left( f_i^{(\mathrm{eq})} \left( \rho, \mathbf u, \mathbf B \right), \vec g_k^{(\mathrm{eq})} \left( \rho, \mathbf u, \mathbf B \right) \right)$ (Eq. \ref{feq})
		\item Compute $s$ and $s^{(\mathrm{eq})}$ (Eq. \ref{contributionExample}, \ref{stressContribution})
		\item Compute $\Delta s_i = s_i - \eq s_i$
		\item Compute $\Delta h_i = h_i - \eq h_i = f_i - \eq f_i - \Delta s_i$
		\item Evaluate $\gamma^{*}$ (Eq. \ref{gammaSolved})
		\item Relax (Collide): $f_i^{'}$ (Eq. \ref{postcollisiongamma}),  and corresponding $\vec g_k^{'}$.
	\end{enumerate}
	
	Standard LB-MHD is recovered when the entropy parameter  is constant and equal to 2 :   $\gamma^*(\vec x,t) = const. = 2$.  However the effect of working with the maximal entropy state for the particle distribution function $f_i$ will have direct effects on the evolution of the magnetic field distribution $\vec g_i$ due to the coupling of the $\vec B$-field in the relaxation distribution function $\eq f$ as well as the coupling of the fluid velocity $\vec u$ in $\eq{\vec g}$.

	\section{Partially Entropic LB-MHD Simulation of the Orszag-Tang Vortex \cite{Orszag}}\label{Sec:Simulation}
	
	Since standard MRT-LB-MHD is recovered when the entropic parameter $\gamma^*(\vec x,t) = 2=const.$ one can readily see the effect of partial entropic stabilization in the variation of $\gamma^*$ from 2.0.  We consider the Orszag-Tang vortex \cite{Orszag} and qualitatively compare our entropic LB simulations with Ref. \cite{Orszag}.
	
\begin{figure}
	\centering
	\begin{tabular}{cc}
		\includegraphics[width=3.25in, height=2.5in, keepaspectratio=true]{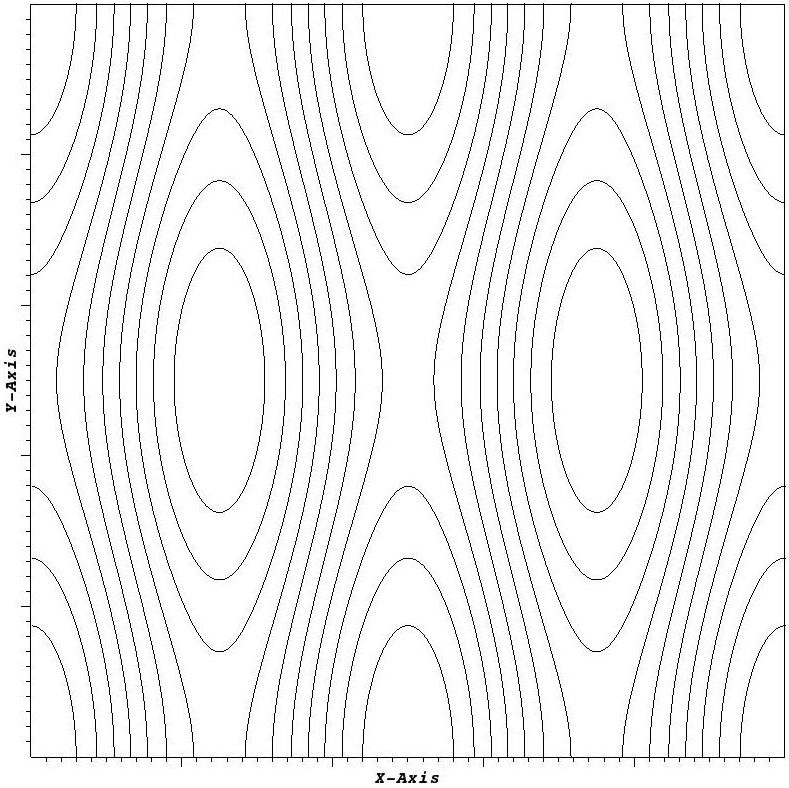}
		&
		\includegraphics[width=2.59in, height=2.5in, keepaspectratio=true]{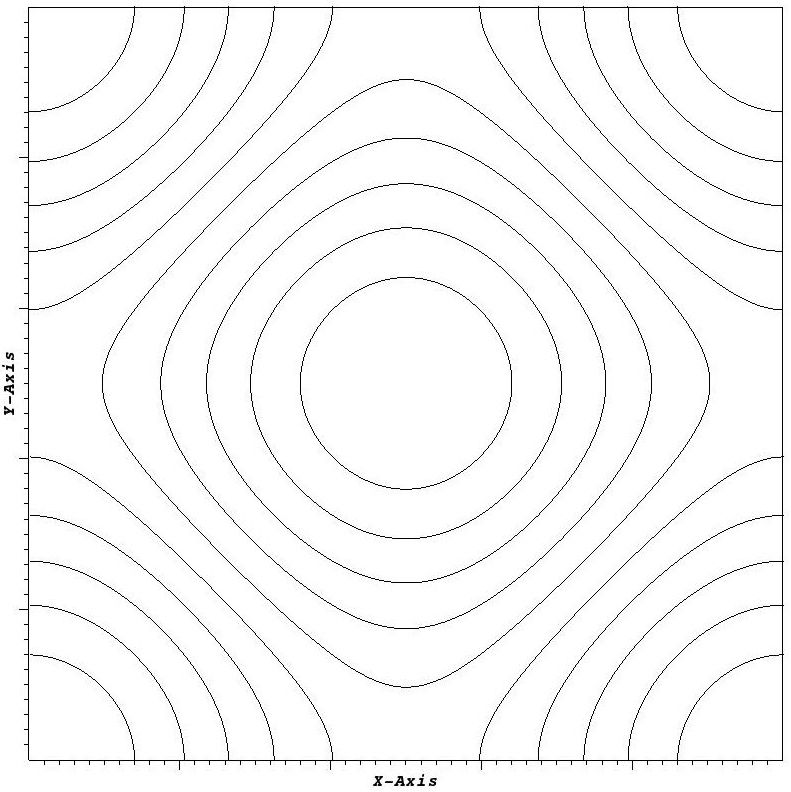}
	\end{tabular}
	\caption[.]{\label{Init}Initial profile of the (a) current and (b) vorticity contours respectively from the partial entropic LB-MHD code.}	
\end{figure}

\begin{figure}
	\centering
	\begin{tabular}{cc}
	\includegraphics[width=3.25in, height=2.5in, keepaspectratio=true]{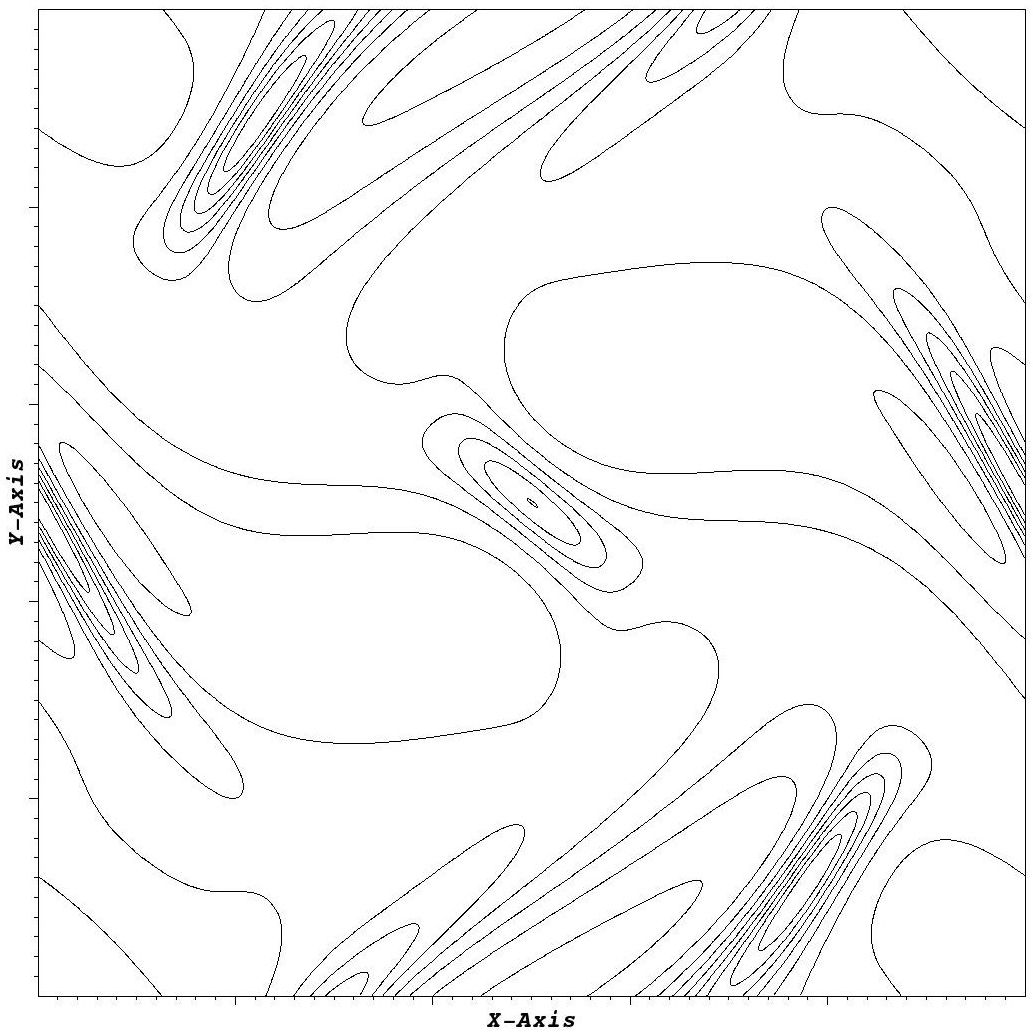}
	&
	\includegraphics[width=2.59in, height=2.5in, keepaspectratio=true]{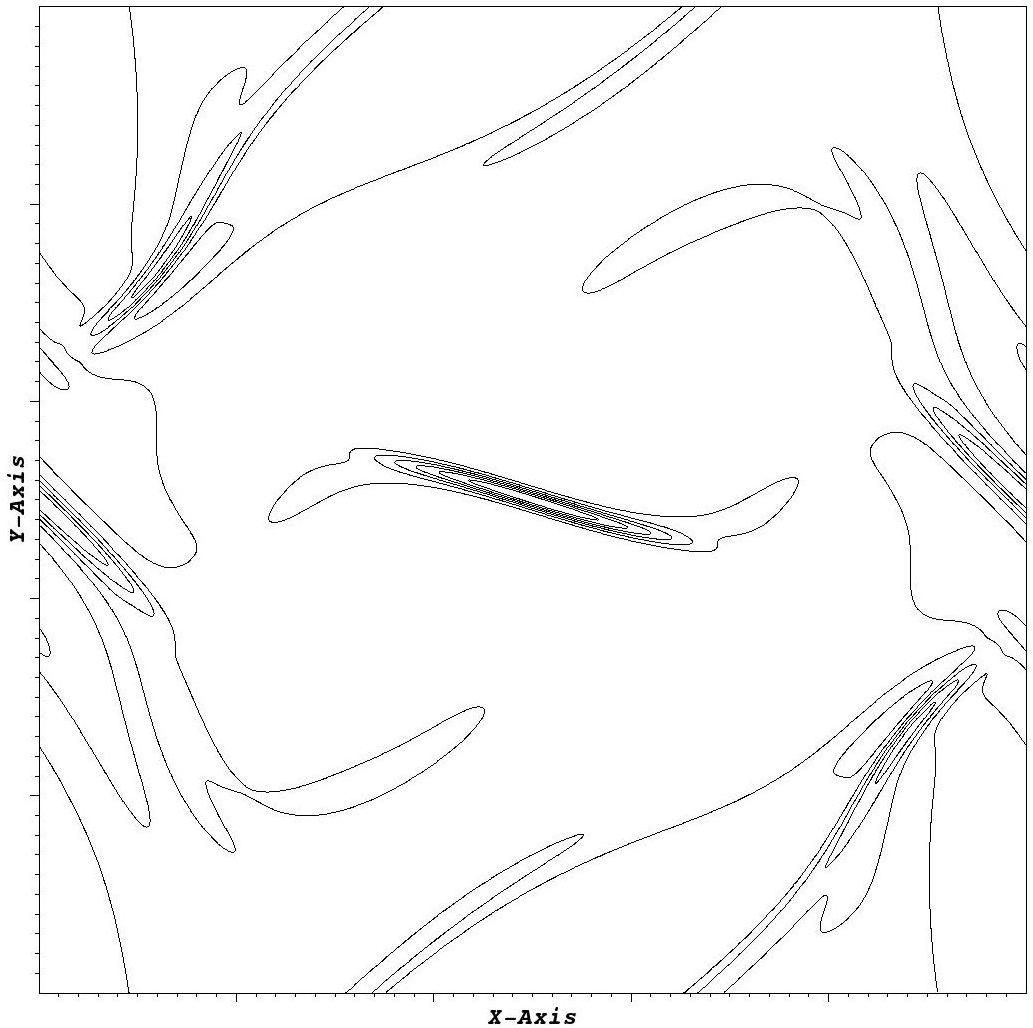}
	\end{tabular}
	\caption[.]{\label{CurrPlot} Snapshot of the current contours from the partial entropic LB-MHD code on a grid of $1024^2$ with $\nu = \eta = 0.005$ (a) at time = 26k, (b) at time = 52k (to be compared with Orszag-Tang\cite{Orszag}, Fig. 7)}	
	\end{figure}

\begin{figure}
	\centering
	\begin{tabular}{cc}
		\includegraphics[width=3.25in, height=2.5in, keepaspectratio=true]{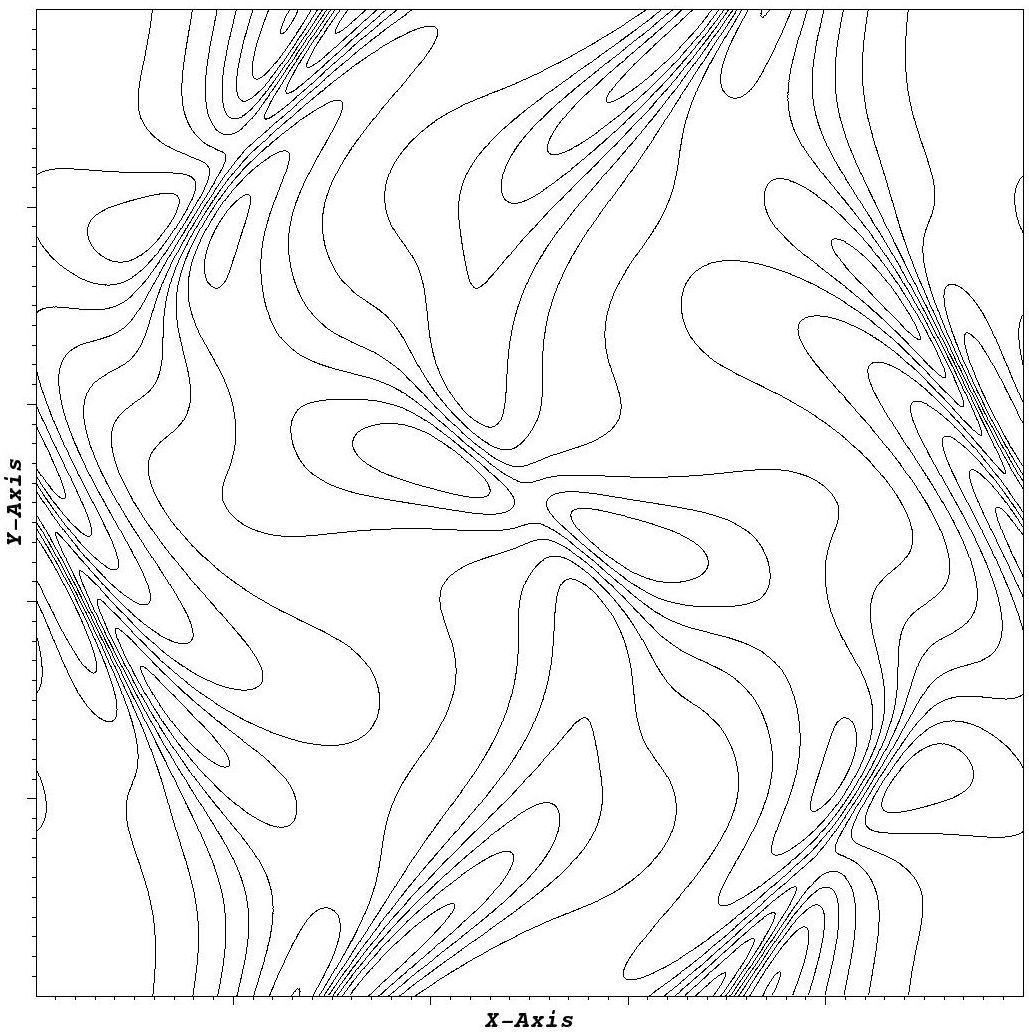}
		&
		\includegraphics[width=2.59in, height=2.5in, keepaspectratio=true]{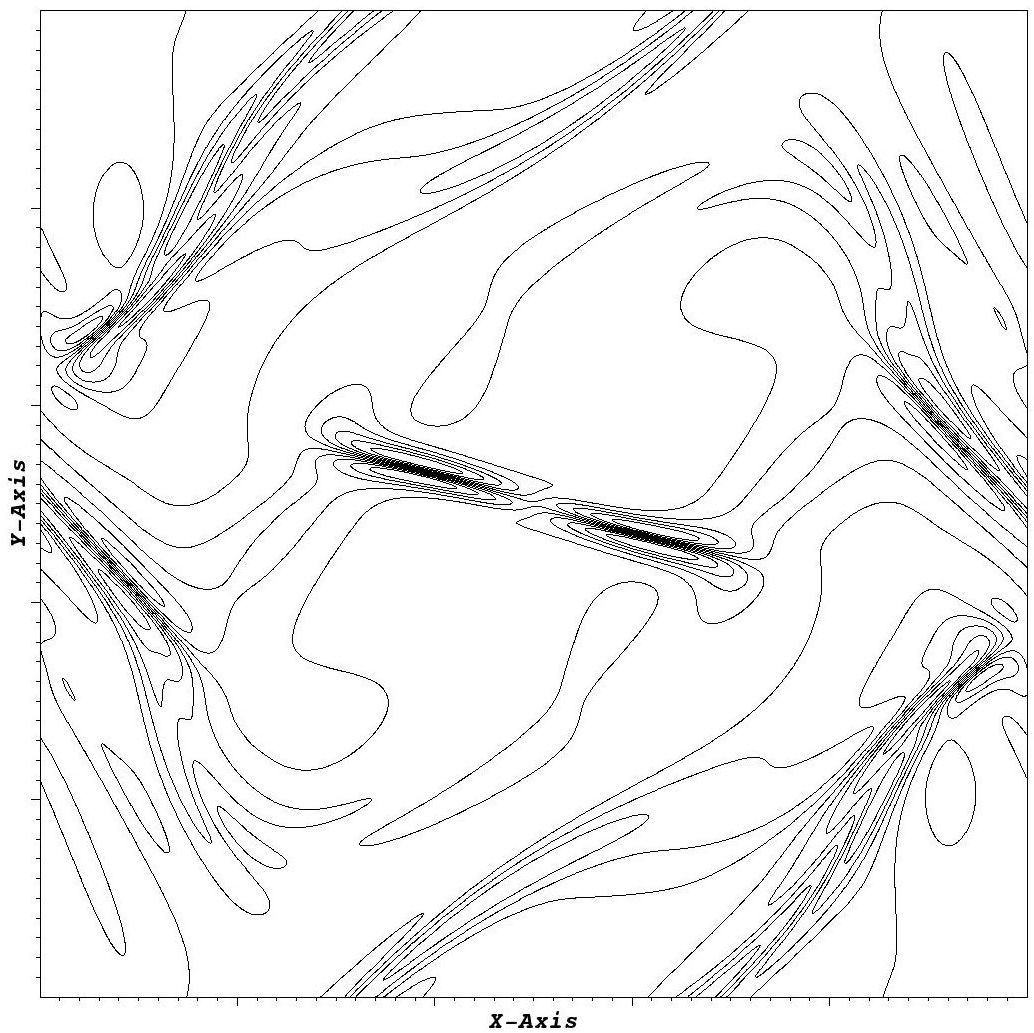}
	\end{tabular}
	\caption[.]{\label{VortPlot}Snapshot of the vorticity contours from the partial entropic LB-MHD code on a grid of $1024^2$ with $\nu = \eta = 0.005$ (a) at time = 26k, (b) at time = 52k (to be compared with Orszag-Tang\cite{Orszag}, Fig. 8)}
	\end{figure}

	Here we show the physics recovered by the variations in the partially entropic parameter $\gamma^{*}$  and its variations away from the MRT value of $\gamma^{*}(\vec x,t) \equiv 2.0$ for the Orszag-Tang vortex.  We qualitatively compare our simulations with the Orszag-Tang\cite{Orszag} profile.
Consider the following initial profiles, Fig. \ref{Init},
	\begin{eqnarray}
	\vec u(x,y,t=0) = U_0 \left[ \sin(y) \hat x  - \sin(x) \hat y \right] \\
	\vec B(x,y,t=0) = B_0 \left[ \sin(y) \hat x  - \sin(2x) \hat y \right]
	\end{eqnarray}
	with $U_0 = B_0 = 6.1 \times 10^{-3}$ and $\nu = \eta = 0.005$.  These correspond to the profiles in \cite{Orszag} with viscosity $\nu$, resistivity $\eta$ and Reynold's number $1250$.
	Snapshots of the current are shown in Fig. \ref{CurrPlot} - these are to be compared with Fig.7 of \cite{Orszag}. Similarly, snapshots of the vorticity are shown in Fig. \ref{VortPlot} - and should be compared to those of Fig 8 in \cite{Orszag}.  One finds excellent qualitative agreement with \cite{Orszag}.

\begin{figure}
	\centering
		\begin{tabular}{cc}
		\includegraphics[width=3.25in, height=2.5in, keepaspectratio=true]{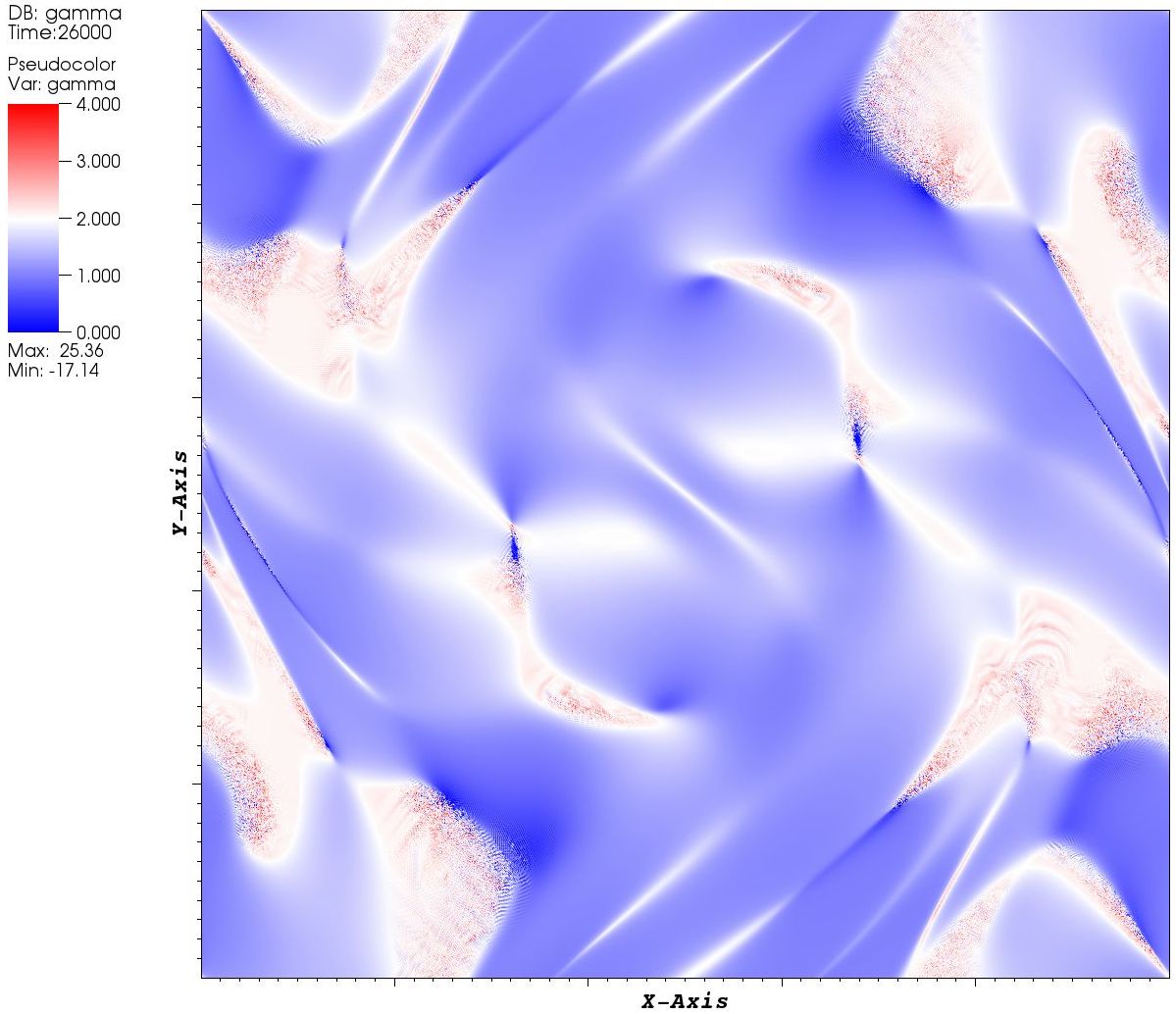}
		&
		\includegraphics[width=3.25in, height=2.5in, keepaspectratio=true]{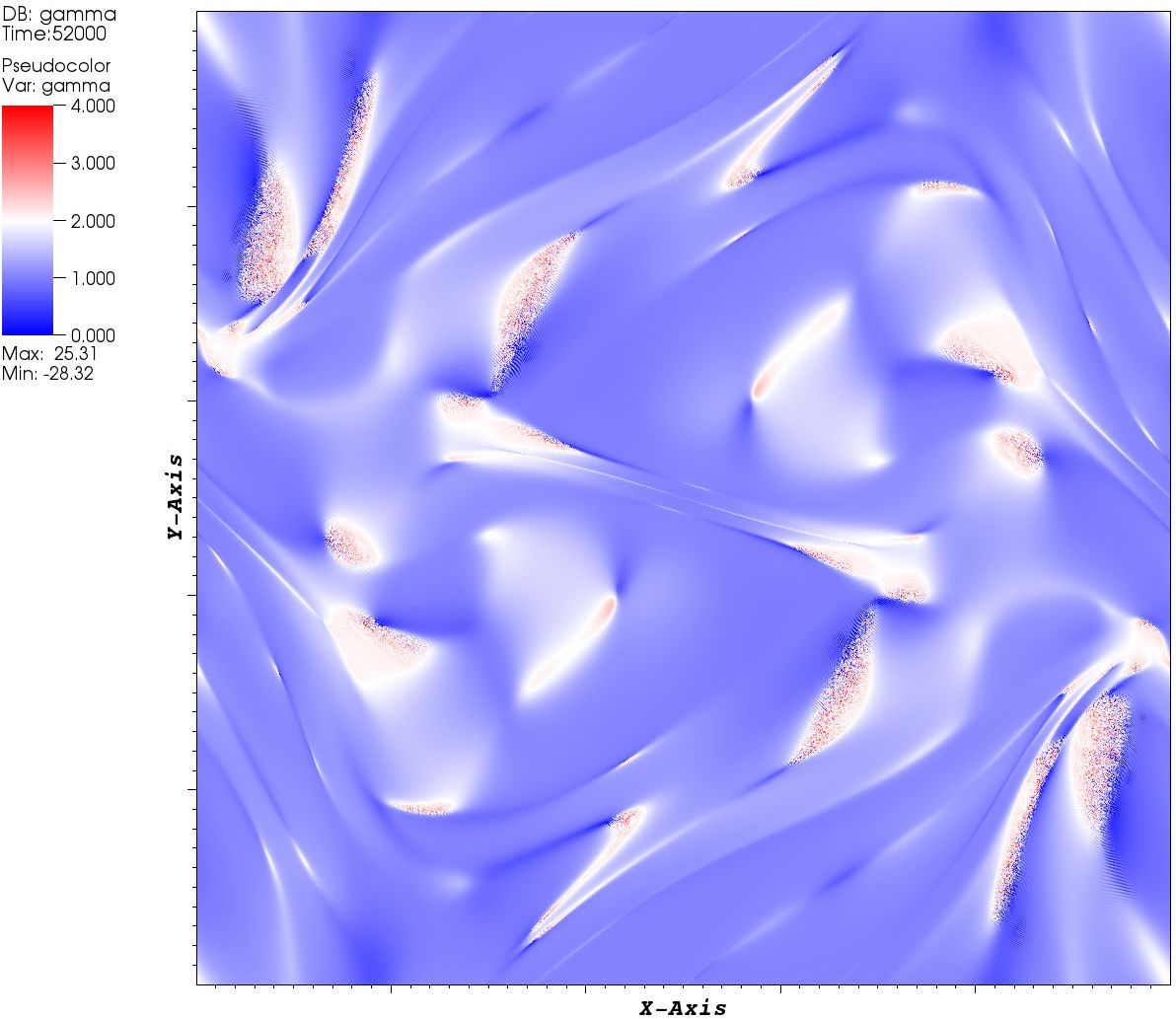}
	\end{tabular}
	\caption[.]{\label{GammaPlot}Plot of the entropic parameter $\gamma^*(x,y)$ after (a) 26k timesteps and (b) 52k timesteps on a $1024^2$ grid with $\nu = \eta = 0.005$.  $\gamma^* = 2.0$ corresponds to ordinary LB-MHD.  Lattice points with $\gamma^* \neq 2.0$ correspond to the stabilizing effects of the partial entropic LB-MHD algorithm.}
	\end{figure}

\begin{figure}
	\centering
		\includegraphics[width=3.25in, height=2.5in, keepaspectratio=true]{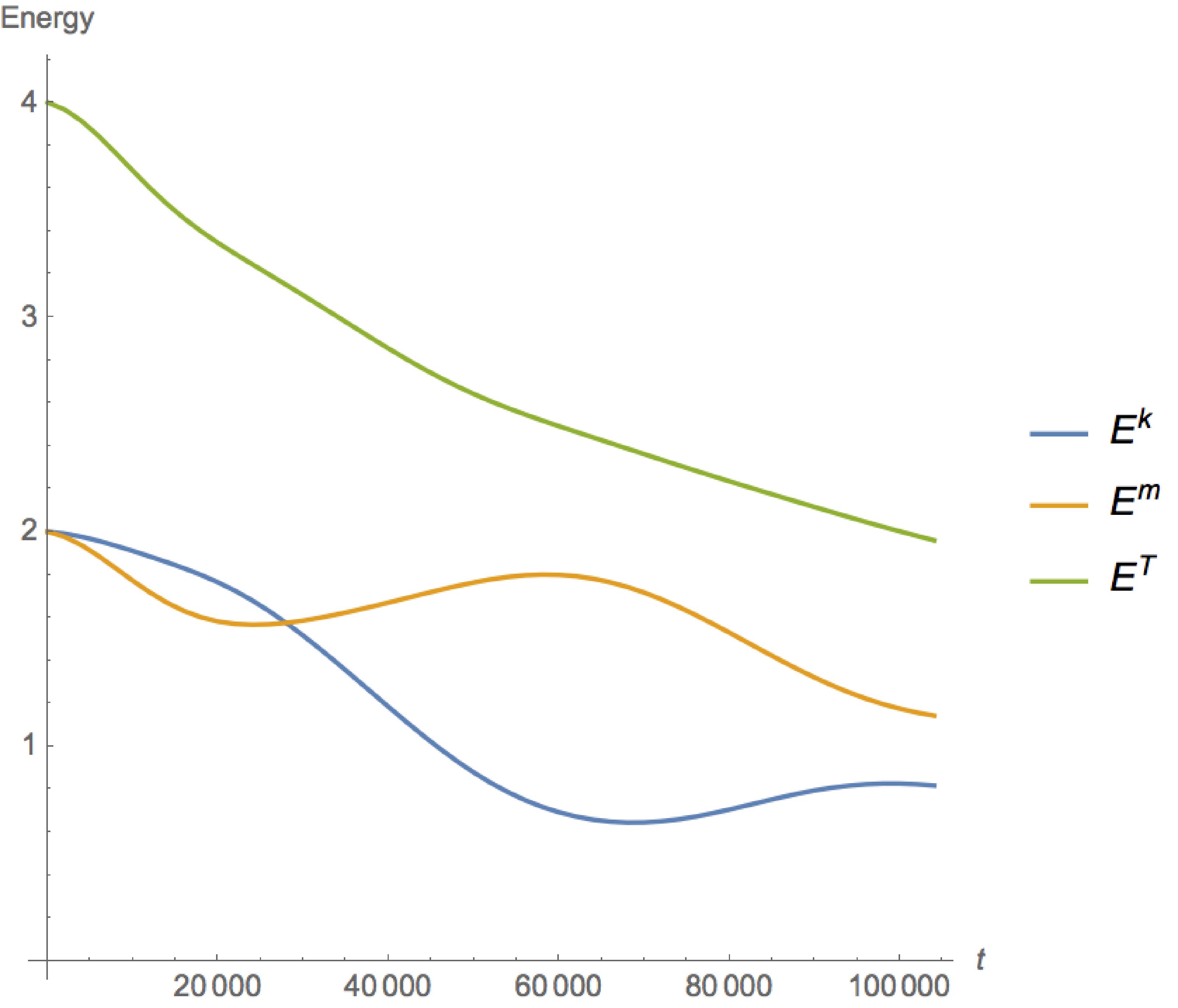}
	\caption[.]{\label{EnergyPlot}Plot of kinetic $(E^k)$, magnetic $(E^m)$, and total $(E^T)$ energies as a function of timesteps $t$ for the partial entropic LB-MHD code on a grid of $1024^2$. $\nu = \eta = 0.02$ for comparison against Orszag and Tang's \cite{Orszag} Fig. 5.}
\end{figure}

	In Fig. \ref{GammaPlot} we plot the corresponding 2D entropy parameter $\gamma^{*}(x,y)$ at these two time snapshot, $t = 26K$ and $t = 52K$ iterations.  The lattice points at which  $\gamma^{*}(x,y) \neq 2$ correspond to spatial positions in which there are effects of our partial entropic LB-MHD algorithm. The energy over time for the Orszag-Tang vortex at $\nu = \eta = 0.02$ is shown in fig. \ref{EnergyPlot} for comparison against Fig. 5 in \cite{Orszag}.
	
	\section{Conclusion}\label{Sec:Conclusion}

	We have extended the Karlin  \cite{entropic1,entropic2} entropic Navier-Stokes algorithm to LB-MHD and tested the ensuing model on an Orszag-Tang vortex.  We considered the D2Q9 model for both the particle and vector magnetic field distributions.  The partial entropy algorithm is applied only to the particle distributions while in using a vector distribution for the magnetic field we are automatically enforcing the $\div \vec B = 0$ constraint while permitting magnetic field reversals.  The algorithm extends immediately to 3D, but because of the much greater computational costs we have restricted our simulations to 2D.  In 2D MHD one can still capture turbulence and the generation of small scale motions since in 2D MHD energy cascades to small scales.  We have found good agreement with the CFD simulations of Orszag and Tang \cite{Orszag}.  Moreover the extreme parallelization of this partial entropic LB-MHD algorithm is retained since this algebraic entropic parameter $\gamma^*$ is determined purely from local information at each lattice site. The accuracy of the under-resolved Navier-Stokes simulations of B\"{o}sch \textit{et. al.} \cite{Bosch} portend that this new (partial) entropy method could be a possible subgrid model in itself.  This partial entropic LB-MHD algorithm is a subset of MRT models in which there is now a dynamical relaxation rate determined for quasi-stabilization of the fluid flow by a well-defined procedure as opposed to the standard static MRT relaxation rates.	
	\section*{Acknowledgments}
	This work was supported by an NSF grant 131424.  The computations were performed on DoD Supercomputer \emph{Topaz}.
	
	\bibliographystyle{unsrt}
	\bibliography{REntropicBib}
	
\end{document}